\documentclass[11pt,a4paper]{article}

\author{Fr\'ed\'eric Bucci$^{1,6}$, Michael Benzaquen$^{2,3,6}$, \\
Fabrizio Lillo$^4$, Jean-Philippe Bouchaud$^{3,5,6}$\\}
\date{\small%
$^1$ \textit{Scuola Normale Superiore, Piazza dei Cavalieri 7, 56126 Pisa, Italy} \\%
$^2$ \textit{Ladhyx  UMR CNRS 7646 \& Department of Economics, Ecole polytechnique,\\ 91128 Palaiseau Cedex, France} \\%
$^3$ \textit{Capital Fund Management, 23-25, Rue de l'Universit\'e 75007 Paris, France}  \\%
$^4$ \textit{Department of Mathematics, University of Bologna,  Piazza di Porta San Donato 5,\\ 40126 Bologna, Italy}\\%
$^5$ \textit{CFM-Imperial Institute of Quantitative Finance, Department of Mathematics, Imperial College, 180 Queen's Gate, London SW7 2RH} \\[2ex]%
$^6$ \textit{Chair of Econophysics and Complex Systems, Ecole polytechnique, 91128 Palaiseau Cedex, France}\\%
\today}
\title{Slow decay of impact in equity markets: \\insights from the ANcerno database}

\usepackage{amssymb}
\usepackage{graphicx}
\usepackage{amsmath}
\usepackage{amsfonts}
\usepackage{geometry}
\usepackage{bm}
\usepackage{breqn}
\usepackage{textcomp}
\usepackage{xcolor}
\usepackage{subfig}
\usepackage{amsmath}
\usepackage{amssymb}
\usepackage{geometry}
\usepackage{mathtools}
\usepackage{bbm}
\usepackage{tabularx}

\newcommand{\avg}{\mathbb{E}}

\newcommand{\sign}{\mathrm{sign}}

\begin{document}

\maketitle

\abstract{We present an empirical study of price reversion after the executed metaorders. We use a data set with more than 8 million metaorders executed by institutional investors in the US equity market. We show that relaxation takes place as soon as the metaorder ends: while at the end of the same day it is on average $\approx 2/3$ of the peak impact, the decay continues the next days, following a power-law function at short time scales, and apparently converges to a non-zero asymptotic value at long time scales (${\sim 50}$ days) equal to $\approx 1/2$ of the impact at the end of the first day. Due to a significant, multiday correlation of the sign of executed metaorders, a careful deconvolution of the \emph{observed} impact must be performed to extract the estimate of the impact decay of isolated metaorders.}

\tableofcontents
\section{Introduction}

It is now well documented that a series of buy (/sell) trades, originating from the same institution, pushes on average the price upwards (/downwards), by a quantity proportionnal to the {\it square-root} of the total volume $Q$ of the buy trades -- see e.g. \cite{ Almgren, Torre, Engle, Bacry, Gomes, Bershova, Bnp, Bonart,Toth,Frazzini}. What happens when such a series of trades (often called a ``metaorder'') is completed? One expects that once the extra buy (/sell) pressure subsides, impact will revert somewhat. There is however no consensus on the detailed behaviour of such {\it impact decay}. The long-run asymptotic price after the metaorder is expected to depend on the information on which trading is based, so it is customary to decompose the observed impact into a ``reaction'' impact, that is a mechanical property of markets, unrelated to information, and a ``prediction impact'' that depends on the quality of information contained in the trade \cite{TQP}.

{From a modeling point of view, several hypotheses have been put forward.} In the stylized, ``fair pricing'' theory of Farmer, Gerig, Lillo \& Waelbroeck (FGLW) \cite{Farmer}, an equilibrium condition is derived between liquidity providers and a broker aggregating informed metaorders from several funds, in which the average price payed during the execution is equal to the price at the end of the reversion phase. If metaorder size distribution is power law with tail exponent $3/2$, the observed impact is predicted to decay towards a plateau value whose height is 2/3 of the peak impact, i.e. the impact reached exactly when the metaorder execution is completed. Within the ``latent order book'' model \cite{Donier,Toth}, reaction impact is expected to decay as a power-law of time, reaching a (small) asymptotic value after times corresponding to the memory time of the market \cite{Benzaquen}. A similar behaviour is predicted by the propagator model \cite{JPpropa}. Note that while the latent order book model predicts that the permanent reaction impact is linear in $Q$ \cite{Benzaquen} (in agreement with no-arbitrage arguments \cite{Huberman,Rosenbaum}), the FGLW theory implies that permanent impact scales as the peak impact, i.e. as $\sqrt{Q}$.

As far as empirical data is concerned, the situation is rather confusing, {mostly because the determination of the time when the relaxation terminates is not unique}. Some papers, {determining permanent impact shortly after the end of the metaorder}, have reported results compatible with the FGLW 2/3 factor \cite{Zarinelli,Gomes,Bershova,Bnp, Moro}, although Gomes \& Waelbroeck \cite{Gomes} note that the impact of uninformed trades appears to relax to zero. Brokmann et al. \cite{Brokmann}, on the other hand, underline the importance of metaorders split over many successive days, as this may strongly bias upwards the apparent plateau value. After accounting for both metaorder correlations and prediction impact, Brokmann et al. \cite{Brokmann} conclude that impact decays as a power-law over several days, with no clear asymptotic value. The work of Bacry et al. \cite{Bacry} leads to qualitatively similar conclusions. 

In the present paper, we revisit this issue using the same ANcerno database as Zarinelli et al. \cite{Zarinelli}, with a closer focus on impact decay. Extending the time horizon beyond that considered in \cite{Zarinelli}, we establish unambiguously that impact decays below the 2/3 plateau, {which is observed as average value of the impact at the end of the same day of the metaorder execution. Specifically,} we find that the overnight contribution to impact decay is  small, in agreement with the idea that the decay takes place in ``volume time'' rather than in calendar time. After accounting for metaorder temporal correlations, impact decay is well fitted by a power-law for intraday time scales and an exponentially truncated power-law for multiday horizons, extrapolating to a plateau value $\approx 1/3$ of the peak impact beyond several weeks.  For such long time lags, however, market noise becomes dominant and makes it difficult to conclude on the asymptotic value of impact, which is a proxy for the (long time) information content of the trades in our database. 

\section{Data \& Definitions}
\label{ANcerno}

\subsection{The ANcerno Dataset}

We use an heterogeneous dataset provided by ANcerno\footnote{ANcerno Ltd (formerly the Abel Noser Corporation)  is a widely recognized consulting firm that works with institutional investors to monitor their equity trading costs. Its clients include many pension funds and assets managers. Previous academic studies that use ANcerno data to investigate the market impact at different times scales includes \cite{Zarinelli, Bucci, BBLB}. See {\tt www.ancerno.com} for details.}, a leading transaction cost analysis provider.  The structure of this dataset allows the identification of metaorders relative to the trading activity of a diversified pool of (anonymized) institutional investors, although informations relative to the execution style and motive of the transaction are not available.  {We define a metaorder as a series of jointly reported executions of a single investor, through one broker within one day, on a given stock and in a given direction.}
It follows that each metaorder is characterized by a stock symbol, the total volume $Q$ (in number of shares) and the times at the start $t_{\mathrm{s}}$ and at the end $t_{\mathrm{e}}$ of its execution with sign $\epsilon=\pm 1$. Our sample covers for a total of 880 trading days, from January 2007 to June 2010 and we select only stocks in the Russell 3000 index. The cleaning procedure introduced in Ref. \cite{Zarinelli} and used in \cite{Bucci,BBLB} is applied to remove possible spurious effects. In this way the available sample is represented by around 8 million metaorders distributed quite uniformly in time and across market capitalizations representing around  5\% of the total market volume.\footnote{Without the above filters, this number would rise to about 10\%.}

\subsection{Definitions}

To characterize the metaorder execution of $Q$ shares we introduce the following observables as done in Ref. \cite{Zarinelli}: the participation rate $\eta$ and the duration $D$ measured in volume time.  The participation rate $\eta$ is defined as the ratio between the volume $Q$ traded by the metaorder and the whole market volume traded during the execution time interval $[t_{\mathrm{s}},t_{\mathrm{e}}]$ 
\begin{equation}
\eta=\frac{Q}{V(t_{\mathrm{e}})-V(t_{\mathrm{s}})},
\end{equation}
where $V(t)$ is the cumulative volume transacted in the market between the start of the day and time $t$. The metaorder duration in volume time $D$ is expressed as 
\begin{equation}
D = \frac{V(t_{\mathrm{e}})-V(t_{\mathrm{s}})}{V_{\mathrm{d}}},
\end{equation}
where $V_{\mathrm{d}}=V(t_{\mathrm{c}})$ is the total volume traded until the close of the day $t=t_{\mathrm{c}}$. The unsigned daily fraction $\phi:=Q/V_{\mathrm{d}}$ is then given by
\begin{equation}
\phi=\eta \times D.
\end{equation}
The statistics of metaorders' duration, participation rate, number of trades, etc. are detailed in \cite{Zarinelli,Bucci}. For example, the participation rate $\eta$ and the duration $D$ are both well approximated by truncated power-law distributions over several orders of magnitude. 

\subsection{Market Impact}
\label{MI}
Let us briefly recall the main definition of market impact necessary to investigate in the following how the price relaxes after the metaorder execution. An asset manager decides to buy or to sell a metaorder of $Q$ shares sending it at time $t=t_{\mathrm{s}}$ to a broker or to an execution algorithm where it is executed sequentially in smaller orders on market until to completion at time $t=t_{\mathrm{e}}$.  The market impact is usually defined in terms of the rescaled log-price $s(t):=(\textrm{log}~ S(t))/\sigma_{\mathrm{d}}$, where $S(t)$ is the price, $\sigma_{\mathrm{d}}=(S_{\textrm{High}}-S_{\textrm{Low}})/S_{\textrm{Open}}$ is a noisy estimator of the daily volatility, estimated from the daily high, low and open prices.  

Given the rescaled average market mid-price at the start of the metaorder $s(t_{\mathrm{s}})$ and the end of its execution, $s(t_{\mathrm{e}})$, we quantify its ``Start-to-End'' price impact $I_{SE}$ with the following antisymmetric expectation
\begin{equation}\label{eq:impact-def}
I_{SE}(\phi)=\avg[\epsilon \cdot (s(t_{\mathrm{e}}) - s(t_{\mathrm{s}}))|\phi]
\end{equation}
where $\epsilon=\pm 1$ is the signed order size of the metaorder with volume $Q$. In practice, we compute the market impact curve $I_{SE}(\phi)$ by dividing the data into evenly populated bins according to the volume fraction $\phi$ and computing the conditional expectation of impact for each bin \cite{Zarinelli, Bacry, Bucci}. Henceforth, error bars are determined as standard errors. Similarly, we will define the ``Start-to-Close'' impact $I_{SC}$ by replacing in Eq. (\ref{eq:impact-def}) the end price $s(t_{\mathrm{e}})$ by the close (log)price of the day $s(t_{\mathrm{c}})$.  

\section{Results}

\subsection{Intraday impact and post-trade reversion}

\begin{figure}
\begin{minipage}{1.0\textwidth}
\centering
\includegraphics[width=0.9\linewidth]{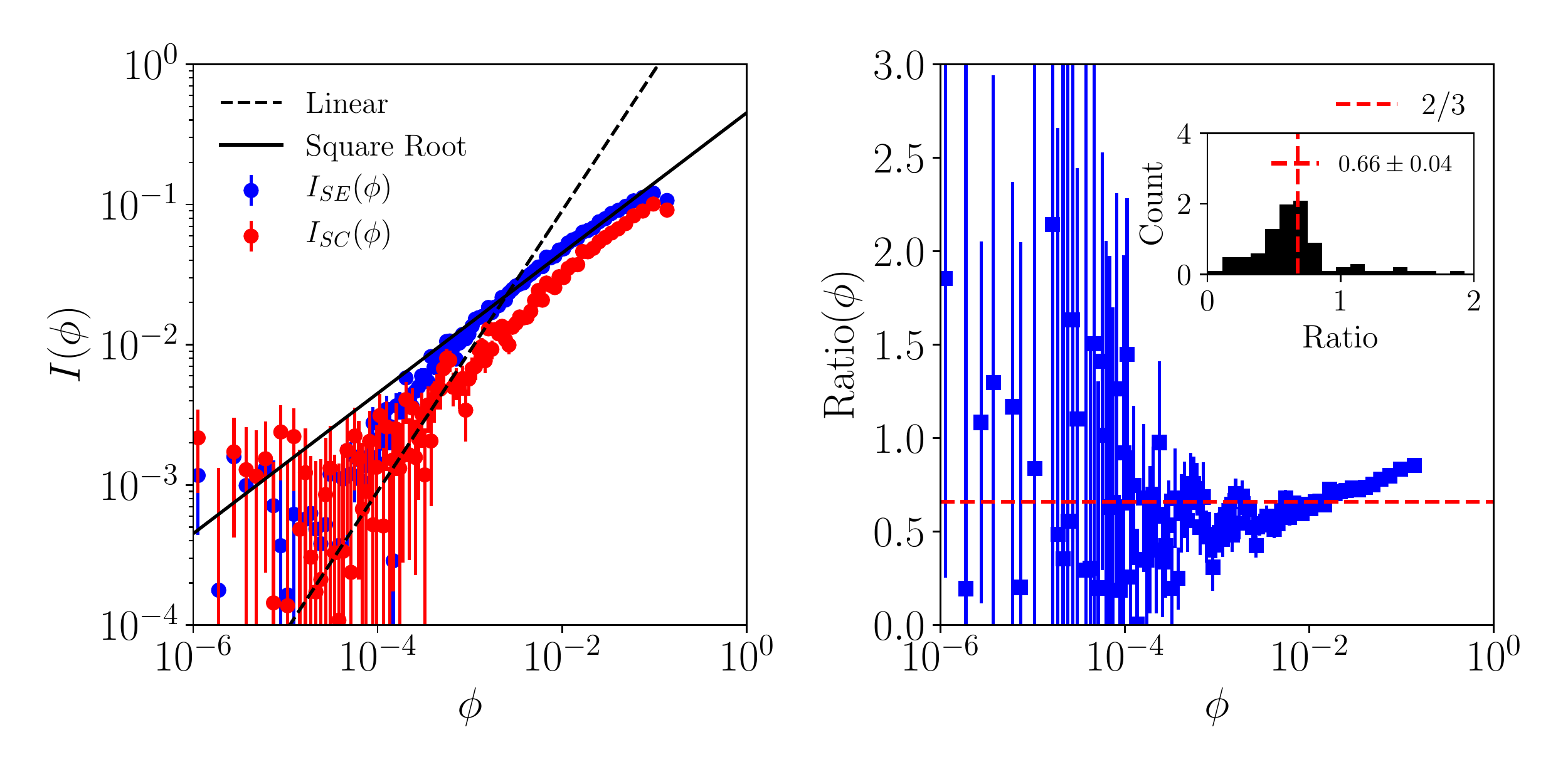}
\end{minipage}
\caption{(Left panel) Start-to-End impact $I_{SE}$ and Start-to-Close impact $I_{SC}$ as a function of the daily volume fraction $\phi$. We also show the square-root impact law $I \propto \sqrt{\phi}$ (plain line) and a linear impact law (dotted line). The slope of $I_{SC}$ appears to be larger than that of $I_{SE}$ as a consequence of a stronger impact decay contribution for smaller $\phi$'s. (Right panel) The ratio $I_{SC}(\phi)/I_{SE}(\phi)$, computed in each volume fraction bin $\phi$. Its average over all $\phi$ is $=0.66 \pm 0.04$. The empirical distribution of the ratio is presented in the inset. Note that for $\phi \gtrsim 10^{-3}$, this ratio increases with $\phi$.}
\label{figESEC}
\end{figure}

In Fig.~\ref{figESEC}, we show the Start-to-End impact $I_{SE}$ and Start-to-Close impact $I_{SC}$ as a function of the daily volume fraction $\phi$. Clearly, $I_{SE}$ behaves as a square-root of the volume fraction $\phi$ in an intermediate regime $10^{-3} \lesssim \phi \lesssim 10^{-1}$, as reported in many previous studies~\cite{Almgren,Torre,   Engle, Bacry, Gomes, Bershova, Bnp, Bonart, Toth,Frazzini}. For smaller volume fractions, impact is closer to linear \cite{Zarinelli} -- see \cite{BBLB} for a recent discussion of this effect. The Start-to-Close impact, measured using exactly the same metaorders, is below the Start-to-End impact ($I_{SC} < I_{SE}$), showing that some post-trade reversion has taken place between the metaorder completion time $t_{\mathrm{e}}$ and the market close time $t_{\mathrm{c}}$. 

The {\it ratio} between these two impact curves is plotted in Fig.~\ref{figESEC} (Right panel). Interestingly, the mean value over all $\phi$ is found to be $0.66 \pm 0.04$, is close agreement with the 2/3 ratio predicted by FGLW, thus confirming previous empirical findings \cite{Zarinelli, Gomes,  Bershova, Bnp, Moro}. However, a closer look at the plot reveals that the ratio systematically increases as $\phi$ increases. Since larger metaorders (i.e. large $\phi$) tend to take longer to execute, one expects the End-to-Close time $T_{EC} = t_{\mathrm{c}} - t_{\mathrm{e}}$ to decrease as $\phi$ increases. Therefore impact decay between the end of the metaorder and the end of the day should be, on average, more effective for small $\phi$.

In order to validate this hypothesis, we now characterise the intraday price reversion by computing the ratio $I_{SC}/I_{SE}$ as a function of the variable $z=V_{EC}/V_{SE}$, where $V_{EC}=V(t_{\mathrm{c}})-V(t_{\mathrm{e}})$ and $V_{SE}=V(t_{\mathrm{e}})-V(t_{\mathrm{s}})$ are respectively the total market volume executed in the time intervals $T_{EC}$ and $T_{SE}$ (similar results -- not shown -- are obtained as a function of $z'=T_{EC}/T$, where $T=t_{\mathrm{e}} - t_{\mathrm{s}}$ is the metaorder execution time). The results are shown in Fig.~\ref{fig5} (Left panel). One clearly sees that impact decays continuously as $z$ increases, and is in fact well fitted by the prediction of the propagator model \cite{TQP,JPpropa}, namely $\mathcal{I}_{\text{prop}}(z)=(1+z)^{1-\beta}-z^{1-\beta}$ with $\beta=0.22$.\footnote{To note, $\beta$ is the decay exponent of the propagator, $G(t)\sim t^{-\beta}$, see e.g. \cite{TQP}.} If one restricts to a smaller interval $z \in [0,2]$, as in \cite{Zarinelli, Bnp}, one finds that the decay appears to saturate around the $2/3$ value (see Fig.~\ref{fig5}, Inset), but zooming out leaves no doubt that impact is in fact decaying to smaller values.

\subsection{Next day reversion}

Quite interestingly, impact decays much in the same way over the next day: in the same figure we plot $I_{SC_2}/I_{SE}$ as a function of $z=V_{EC_2}/V_{SE}$, where $C_2$ refers to the close of the next day and $V_{EC_2}=V_{EC}+ V_{\mathrm{d}}$ (i.e. the overnight does not contribute to $T_{EC_2}$). Provided one applies a factor $\zeta \approx 4/5$ that accounts for the autocorrelation of metaorders (see next section, and Fig. \ref{fig10} (Left panel))\footnote{More precisely, we have set $\zeta=1/(1+C(1))$, where $C(\tau)$ is the autocorrelation function plotted in 
Fig. \ref{fig10} (Left panel).}, the next day impact decay nicely falls in the continuation of the intraday decay, and is also well accounted for by the very same scaling function $\mathcal{I}_{\text{prop}}(z)$. 

\begin{figure}
\begin{minipage}{1.0\textwidth}
\centering
\includegraphics[width=1.0\linewidth]{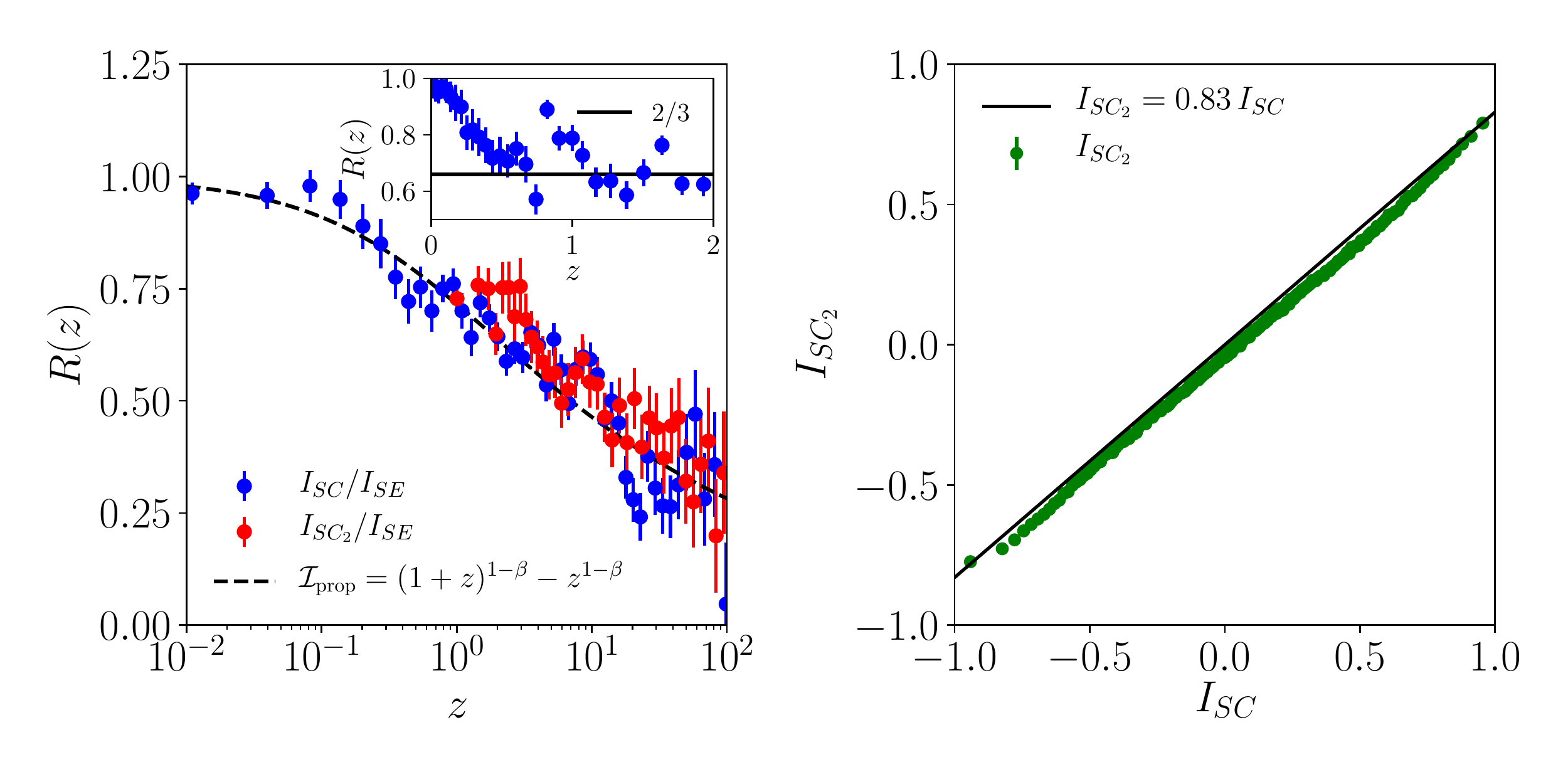}
\end{minipage}
\caption{(Left panel) Price relaxation $R(z)$ over two consecutive days. Blue points: impact decay within the same day of the metaorder's execution, i.e. $I_{SC}/I_{SE}$ as a function of $z=V_{EC}/V_{SE}$, in a semi-log scale. Red points: impact decay using the close of the next day, i.e. $\zeta I_{SC_2}/I_{SE}$ as a function of $z=V_{EC_2}/V_{SE}$, with $\zeta=0.80$. Both sets of points are well fitted by the prediction of the propagator model: $\mathcal{I}_{\text{prop}}(z)=(1+z)^{1-\beta}-z^{1-\beta}$ with $\beta=0.22$. (Inset): Same day impact decay, in a lin-lin plot restricted to $z \in [0,2]$, suggesting relaxation towards a $2/3$ value (horizontal line). (Right panel) Average of Start-to-next day Close $I_{SC_2}$, conditioned to different values of $I_{SC}$. The regression lines yield $I_{SC_2} = 0.83 \, I_{SC}$.}
\label{fig5}
\end{figure}

Fig.~\ref{fig5} (Right panel) provides complementary information: we show the average of $I_{SC_2}$ conditioned to different values of $I_{SC}$, which clearly demonstrates that these two quantities are proportional and related to the same decay mechanism. It shows in particular that the impact measured at the end of the next day would still behave as a square-root of the volume of the executed metaorder. 

\subsection{Impact decay over multiple days}

\begin{figure}[t!]
\begin{minipage}{1.0\textwidth}
\centering
\includegraphics[width=1.0\linewidth]{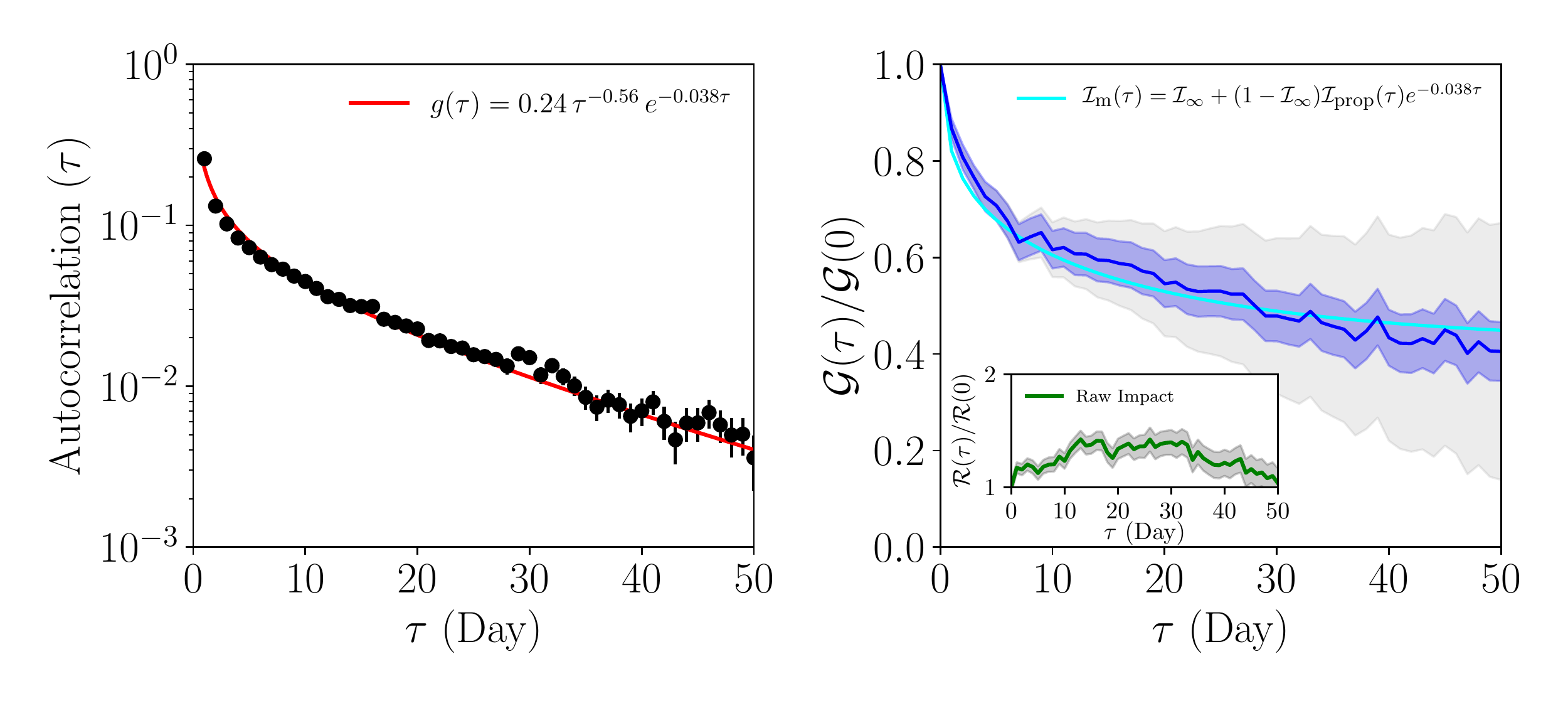}
\end{minipage}
\caption{(Left panel) Empirical autocorrelation of the signed square-root volume imbalance $\Phi^{\bullet 1/2}$, as a function of the lag $\tau$, averaged over all stocks. This autocorrelation persists over many days, as is fitted as an exponentially truncated power law $g(\tau) = a \tau^{-\gamma} e^{-b \tau}$ with $a=0.24 \pm 0.04$ and $b=0.038 \pm 0.002$ (corresponding to $1/d\simeq 26$ days). The value of the exponent $\gamma$ is fixed to $1 - 2 \beta = 0.56$, as dictated by the propagator model \cite{JPpropa}. (Right panel) Normalized decay kernel $\mathcal{G}(\tau)/\mathcal{G}(0)$ estimated using Eq. \ref{Itau} for $\tau \in [1,50]$ days. The fit corresponds to the exponentially truncated modified propagator model  $\mathcal{I}_{\text{m}}(\tau)$ with $b=0.038$ (see Left panel), which provides an asymptotic decay level $\mathcal{I}_\infty \approx 0.42 \pm 0.01$ (fit errors only). The error bars on the graph are (i) bootstrap errors (blue region) and (ii) cumulated regression errors (grey region). Inset: Normalized ``bare'' response function $\mathcal{R}(\tau)/\mathcal{R}(0)$ as a function of $\tau$ (see definition in Eq. (\ref{Rtau})).}
\label{fig10}
\end{figure}

Having established that impact decay occurs both intraday and during the next day, it is tempting to conjecture that impact will continue to decay on longer time scales. However, the empirical investigation of such a decay faces several hurdles. First, as the time lag increases, the amount of noise induced by overall market moves becomes larger and larger (in fact as $\sigma_{\mathrm{d}} \sqrt{\tau}$, where $\tau$ is the number of days). Second, metaorders are often executed over several days, leading to long range autocorrelations of the order flow. {This effect, investigated considering metaorders from the same fund \cite{TQP}, is here investigated by considering a very heterogeneous set of funds and illustrated in the left panel of Fig. \ref{fig10}. We find that metaorder signs autocorrelation is well fitted by an exponentially truncated power law with a time scale of $\approx 26$ days} and an exponent $\gamma$ fixed by the propagator model constraint $\gamma=1-2 \beta$. Intuitively, these correlations may mask the decay of impact, as trades in the same direction during the following days tend to counterbalance impact reversion, leading to an apparent increase of impact (see Fig. \ref{fig10}, Right panel, Inset). This contribution should be somehow removed to estimate the ``natural'' decay of impact. 

A possible way to overcome the latter problem is to apply a deconvolution method introduced by Brokmann et al. \cite{Brokmann}. We will study our metaorder database at the daily time scale with the same angle: for each day $\tau=\{1,\cdots,880 \}$ and asset we computed the net daily traded volume $\Phi(\tau)=\sum_{i=1}^N \epsilon_i \phi_i$ where $N$ is the number of metaorders in the database, for a given asset and a given day $\tau$. As recently shown in Ref. \cite{Bucci}, the impact of a set of different metaorders, all executed the same day, is well described by an extended square-root law where all metaorders are bundled together:
\begin{equation}
I = Y \times \sigma_{\mathrm{d}} \Phi^{\bullet 1/2},
\end{equation}
where we use the notation $x^{\bullet 1/2} = \sign(x) \sqrt{|x|}$.

The return of the asset between the last day close and the close of each day $\tau$ is denoted $r(\tau)$. The method used in \cite{Brokmann} amounts to assuming a quasi-linear model, i.e.
\begin{equation}
r(\tau)= \beta_{\text{capm}}(\tau) \, r_{\textrm{M}}(\tau) + \sum_{\ell=0}^H G(\ell) \cdot \sigma_{\mathrm{d}} \times \widetilde\Phi^{\bullet 1/2}(\tau-\ell) + \xi(\tau),
\label{regre}
\end{equation}
where $G(\ell)$ are coefficients, $H$ is a certain horizon (taken to be $H=50$ days), $\xi$ is a noise term and $\beta_{\text{capm}}(\tau) \, r_{\textrm{M}}(\tau)$ is the systematic component that takes into account the market drift: $\beta_{\text{capm}}(\tau)$ is the beta of the traded stock computed on the period from $\tau - 20$ to $\tau+20$ and $r_{\textrm{M}}(\tau)$ the daily close-close return of the market (here the Russell 3000 index). Finally, $\widetilde\Phi^{\bullet 1/2}(\tau) = \Phi^{\bullet 1/2}(\tau) - \beta_{\text{capm}}(\tau) \langle \Phi^{\bullet 1/2}(\tau) \rangle_{\text{stocks}}$, where we subtract $\beta_{\text{capm}}$ times the cross-sectional average of the expected impact.

\begin{figure}[t!]
\begin{minipage}{1.0\textwidth}
\centering
\includegraphics[width=1.0\linewidth]{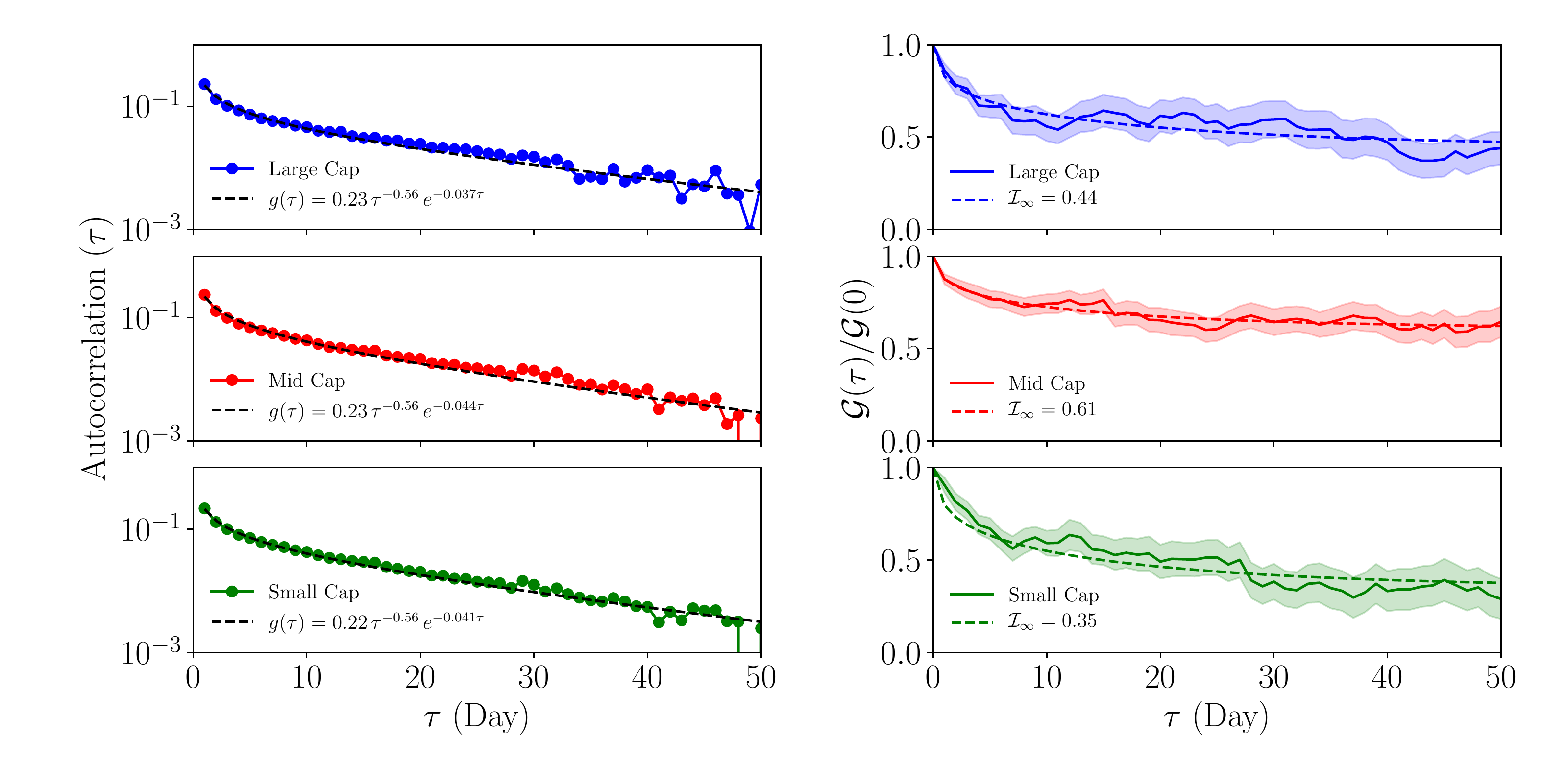}
\end{minipage}
\caption{(Left panel) Empirical autocorrelation of the signed square-root volume imbalance $\Phi^{\bullet 1/2}$, as a function of the lag $\tau$, averaged over all stocks in a given market cap tranche. Each function is fitted as an exponentially truncated power law $g(\tau) = a \tau^{-\gamma} e^{-b \tau}$ with $\gamma = 0.56$. The parameters $a$ and $b$ are very close in the three cases. (Right panel) Normalized decay kernel $\mathcal{G}(\tau)/\mathcal{G}(0)$ estimated using Eq. \ref{Itau} for $\tau \in [1,50]$ days, again using all stocks in a given market cap tranche. The fit corresponds to the exponentially truncated modified propagator model $\mathcal{I}_{\text{m}}(\tau)$ with $d=0.038$, which provides an asymptotic decay level $\mathcal{I}_\infty \approx 0.44$ (large caps), $\mathcal{I}_\infty \approx 0.61$ (mid cap) and $\mathcal{I}_\infty \approx 0.35$ (small cap), all within the grey region of Fig. \ref{fig10}, Right panel.}
\label{fig11}
\end{figure}

{Pooling all the stocks together,\footnote{We have checked that different subsamples of the full sample lead to similar results (for example, slicing the pool of stocks according to their market capitalisation, see Fig. \ref{fig11}).}} a least-square regression allows us to determine the coefficients $G(\ell)$, from which we reconstruct the `reactional' impact kernel $\mathcal{G}(\tau)$ as 
\begin{equation}
\mathcal{G}(\tau)=\sum_{\ell=0}^{\tau} G(\ell).
\label{Itau}
\end{equation}
The kernel $\mathcal{G}(\tau)$ is a proxy of the impact of an isolated metaorder. If the metaorder was uniformed, $\mathcal{G}(\tau)$ would describe the mechanical reaction of 
the market to such a trade. Any non-zero asymptotic value of $\mathcal{G}(\tau \to \infty)$ would either reveal that metaorders are on average informed, or that even random trades have positive permanent impact on prices (as in, e.g. \cite{Benzaquen}).     

{To estimate error bars,} we generated 200 bootstrap samples using all 1500 stocks, and ran the linear regression Eq. (\ref{regre}) on each of them. The average result is shown in Fig. \ref{fig10} (Right panel), together with error bars coming from the least-square regression and from the bootstrap procedure. From this graph, we see that the estimated impact kernel $\mathcal{G}(\tau)$ slowly decays in a time window comparable with the one over which we measure a persistent autocorrelation (as shown in the left panel of Fig. \ref{fig10}). We have fitted the empirically determined, normalized impact kernel $\mathcal{I}_{\text{m}}(\tau) := \mathcal{G}(\tau)/\mathcal{G}(0)$ using an ad-hoc modified propagator kernel, that accounts for a final exponential decay towards an asymptotic value $I_\infty$:
\begin{equation}
\mathcal{I}_{\text{m}}(\tau) =  \mathcal{I}_\infty + (1 - \mathcal{I}_\infty) \mathcal{I}_{\text{prop}}(\tau) e^{-b \tau},
\label{Rtaubis}
\end{equation}
where $b$ is a parameter fixed by the corresponding decay of the flow autocorrelation, see Fig.~\ref{fig10} (Left panel). Keeping the same shape for $\mathcal{I}_{\text{prop}}(\tau)$ as the one describing the short-term decay of impact (i.e. fixing $\beta=0.22$), the one-parameter fit gives $\mathcal{I}_\infty \approx 0.42$. Leaving $b$ free in a 2-parameter fit leads to very similar values: $b=0.03 \pm 0.01$ and $\mathcal{I}_\infty = 0.39 \pm 0.05$. However, setting $b=0$ and leaving $\beta$ free leads to $\beta=0.15 \pm 0.04$ and a zero asymptotic value $\mathcal{I}_\infty = 0.0 \pm 0.19$.

Although the error bars are already large for $\tau=50$, the fit seems to favor a non-zero asymptotic value $\mathcal{I}_\infty \approx 1/2$. Since the impact has on average already decayed to approximately $2/3$ of its peak value at the end of the trading day, this value of $\mathcal{I}_\infty$ suggests a long time asymptotic plateau at $2/3 \times 1/2 \approx 1/2$ of the peak value, significantly below the $2/3$ value predicted by FGLW (see also Fig. \ref{fig5}). This can be taken as a measure of the information content of the trades in the ANcerno database. Since we have no knowledge about the intensity of the trading signal which triggered the metaorders,\footnote{Following Ref. \cite{alpha}, we have attempted to identify ``skilled'' vs ``unskilled'' metaorders. However, since we only have a broker-level identification of the metaorders, this attempt has not been fruitful.} we cannot subtract the ``alpha'' component from the observed returns, as was done in Ref. \cite{Brokmann}, where after removing the alpha of the manager and the contribution of correlated trades, impact was found to decay to $\approx 0.15$ of its initial value after 15 days.  Adding to the regressors of Eq. \ref{regre} past values of $(r - \beta_{\textrm{capm}} r_{\mathrm{M}})$, as a proxy for mean-reversion and/or trending signals that investors commonly use, we find a slightly larger plateau ($\approx 0.54$) when $b$ is kept at the value $0.038$. This reveals how noisy the data is, because one would have expected a decrease of $\mathcal{I}_\infty$ when including more ``alpha'' signal in the regression. However, we do observe mean reversion on short time scales and momentum beyond, as expected. 

Finally, we also show in Fig. \ref{fig10} (Right panel, inset) the full ``response function'' $\mathcal{R}(\tau)$, defined as \cite{JPpropa}:
\begin{equation}
\mathcal{R}(\tau) = \sum_{\tau'=0}^\tau \mathbb{E}[\widetilde r(\tau+\tau') \widetilde \Phi^{\bullet 1/2}(\tau)],
\label{Rtau}
\end{equation}
where $\widetilde r(\tau) := r(\tau) - \beta_{\text{capm}}(\tau) \, r_{\textrm{M}}(\tau)$. This quantity elicits an {\it apparent} evolution of impact, without accounting for metaorder autocorrelations. Such autocorrelations are strong enough to make $\mathcal{R}(\tau)$ increase as a function of $\tau$ (see \cite{TQP, JPpropa} for similar results). This plot illustrates how the autocorrelation of order flow can strongly bias the estimation of impact decay and its asymptotic value (see \cite{Brokmann} for a similar discussion).

\section{Conclusion \& Discussion}
\label{concl}

In this paper we presented an empirical study of the impact relaxation of metaorders executed by institutional investors in the US equity market. We have shown that relaxation takes place as soon as the metaorder ends, and continues the following day with no apparent saturation at the plateau value corresponding to the ``fair pricing'' theory. For example, the impact measured at the next-day close is, on average, around $4/5$ of the impact at the end of the day when the metaorder is executed. 

The decay of impact is described by a power-law function at short time scales, while it appears to converge to a non-zero asymptotic value at long time scales ({$\sim 50$} days), equal to $1/2$ of the initial impact (i.e. at the end of the first day). Due to a significant, multiday correlation of the sign of executed metaorders, a careful deconvolution of the {\it observed} impact must be performed to extract the {\it reaction} impact contribution (where, possibly, some information contribution remains). Once this is done, our results match qualitatively those of Ref. \cite{Brokmann}, obtained on a smaller set of metaorders executed by a single manager (CFM). In particular, we find no support for the prediction of Farmer et al. \cite{Farmer}, that the permanent impact equal to $2/3$ of the peak impact.  

Executing a quantity $Q$ moves the price, on average, as $I(Q) = Y \sigma_{\mathrm{d}} (Q/V_{\mathrm{d}})^{1/2}$, where $Y$ is a certain numerical constant  \cite{ Almgren, Torre, Engle, Bacry, Gomes, Bershova, Bnp, Bonart,Toth,Frazzini}. Assuming that this impact is fully transient and decays back to zero at long times, the corresponding average cost of trading is $2/3 I(Q)$. If the investor predicts a certain price variation $\Delta$, his/her optimal trade size is given by the following maximization problem:
\begin{equation}
Q^* = \text{argmax} \left[ \Delta Q - \frac23  Q \, I(Q) \right] \quad \Rightarrow \quad I(Q^*) = \Delta.
\end{equation}
The last equation means that the investor should trade until his/her average impact pushes the price up to the predicted level $\Delta$, but not beyond. For truly informed investors, there should be no decay of impact at all, since the price has been pushed to its correctly predicted value. For uninformed investors, on the other hand, impact should decay back to zero. Averaging over all metaorders of size $Q$, one should therefore expect an apparent permanent impact given by:
\begin{equation}
\mathcal{I}_\infty(Q) := f(Q) \times I(Q) + (1 - f(Q)) \times I_R(Q),
\end{equation}
where $f(Q)$ is the fraction of metaorders of volume $Q$ that are truly informed and $I_R(Q)$ is the permanent, reactional part of impact -- expected to be zero only if markets were truly efficient. A precise empirical determination of the size dependence of ${I}_R(Q)$ would be extremely interesting. However, this seems to be out of reach: not only would it require a large data set of metaorders reputed to be information-less (such as the portfolio transition trades of Ref. \cite{Gomes}), but also the error on the long-term asymptotic value of $I_R(Q)$ is bound to be very large, as Fig.~\ref{fig10} shows. At this stage, it is thus difficult to confirm or infirm the validity of the theoretical arguments that predict a linear-in-$Q$ dependence of $I_R(Q)$ \cite{Huberman,Rosenbaum,Benzaquen}.

\section*{Acknowledgments}
The authors thank Z. Eisler, A. Fosset, C.-A. Lehalle, I. Mastromatteo, M. Rosenbaum, B. T\'oth and E. Zarinelli for useful discussions.

\section*{Data availability statement}
The data were purchased from the company ANcerno Ltd (formerly the Abel Noser Corporation) which is a widely recognised consulting firm that works with institutional investors to monitor their equity trading costs. Its clients include many pension funds and asset managers. The authors do not have permission to redistribute them, even in aggregate form. Requests for this commercial dataset can be addressed directly to the data vendor.  See www.ancerno.com for details.


\begin{thebibliography}{100}
\bibitem{0}Bouchaud, J.-P., J. D. Farmer, and F. Lillo, 2009, How markets slowly digest changes in supply and demand, \textit{Handbook of Financial Markets: Dynamics and Evolution}, edited by T. Hens and  K. Schenk-Hoppe, pp. 57-160,  Elsevier: Amsterdam.
\bibitem{TQP}Bouchaud, J.-P., J. Bonart, J. Donier, and M. Gould, 2018, Trades, Quotes and Prices: Financial Markets Under The Microscope, \textit{Cambridge University Press}. 
\bibitem{JPpropa}Bouchaud, J.-P., Y. Gefen, M. Potters, and M. Wyart, 2004, Fluctuations and response in financial markets: the subtle nature of random price changes, \textit{Quantitative Finance}, vol. 4, no. 2, pp. 176-190. 
\bibitem{3}Lillo, F., and J. D. Farmer, 2004,  The long memory of the efficient market, \textit{Studies in Nonlinear Dynamics and Econometrics}, vol. 8, no. 3.
 \bibitem{Almgren}Almgren, R.,  C. Thum, E. Hauptmann, and H. Li, 2005,  Direct estimation of equity market impact, \textit{Risk}, vol. 57.
 \bibitem{Torre}Torre, N., 1997, \textit{Barra market impact model handbook,} BARRA Inc., Berkeley.
\bibitem{Engle}Engle, R. F., R. Ferstenberg, and J. Russell, 2008,  Measuring and modeling execution cost and risk, Chicago GSB Research Paper, no. 08-09.
\bibitem{Zarinelli}Zarinelli, E., M. Treccani, J. D. Farmer, and F. Lillo, 2015, Beyond the Square Root: Evidence for Logarithmic Dependence of Market Impact on Size and Participation Rate, \textit{Market Microstructure and Liquidity}, vol. 1, no. 2.
 \bibitem{Huberman} Huberman, G., and W.  Stanzl, 2004, Price manipulation and quasi-arbitrage, \textit{Econometrica}, 72(4), 1247-1275.
 \bibitem{Rosenbaum} Jusselin, P., and M. Rosenbaum, 2018, No-arbitrage implies power-law market impact and rough volatility, working paper. 
 \bibitem{Brokmann}Brokmann, X., E. Serie, J. Kockelkoren, and J.-P. Bouchaud, 2015, Slow decay of
 impact in equity markets, \textit{Market Microstructure and Liquidity}, 1(02), 1550007.
 \bibitem{Bacry}Bacry, E., A. Iuga, M. Lasnier, and C.-A. Lehalle, 2015, Market impacts and the life
 cycle of investors orders, \textit{Market Microstructure and Liquidity}, 1(02), 1550009.
\bibitem{Gomes} Gomes, G., and H. Waelbroeck, 2015, Is Market Impact  a Measure of the Information Value of Trades? Market Response to Liquidity vs Informed Trades, \textit{Quantitative Finance} 15,  773-793.
\bibitem{Bershova} Bershova, N., and D. Rakhlin, 2013, The Non-Linear Market Impact of Large Trades: Evidence from Buy-Side Order Flow, \textit{Quantitative Finance} 13, 1759-1778.
\bibitem{Donier}Donier, J., J. Bonart, I. Mastromatteo, and J.-P. Bouchaud, 2015, A fully consistent, minimal model for non-linear market impact, \textit{Quantitative Finance} 15, 1009-1121.
\bibitem{Bucci} Bucci, F., I. Mastromatteo, Z. Eisler, F. Lillo, J.-P. Bouchaud, and C.-A. Lehalle, 2018, Co-impact: Crowding effects in institutional trading activity, https://arxiv.org/abs/1804.09565.
\bibitem{Farmer} Farmer, J. D., A. Gerig, F. Lillo, and H. Waelbroeck, How efficiency shapes market impact, \textit{Quantitative Finance}, vol. 13, no. 11, pp. 1743-1758.
\bibitem{Benzaquen} Benzaquen, M., and  J.-P. Bouchaud, 2017, Market impact with multi-timescale liquidity, \textit{Quantitative Finance} 11, 1781-1790.
\bibitem{Bnp}Said, E., A. B. H. Ayed, A. Husson, and F. Abergel, 2018, Market Impact: A systematic study of limit orders, \textit{Market Microstructure and Liquidity}, vol. 03, no. 03n04.
\bibitem{Bonart}Donier, J., and J. Bonart, 2015, A Million Metaorder Analysis of Market Impact on the Bitcoin, \textit{Market Microstructure and Liquidity}, vol. 01, no. 02.
\bibitem{Moro} Moro, E., J. Vicente, L. G. Moyano, A. Gerig, J. D. Farmer, G. Vaglica, F. Lillo,  and R. N. Mantegna, 2009, Market impact and trading profile of hidden orders in stock markets. \textit{Physicak Review E}, 80(6), 066102.
 \bibitem{Toth}T\'oth, B., Y. Lemperiere, C. Deremble, J. De Lataillade, J. Kockelkoren, and
J.-P. Bouchaud, 2011, Anomalous price impact and the critical nature of liquidity in financial markets. \textit{Physical Review X}, vol. 1, no. 2, p. 021006.
\bibitem{alpha} Saglam, M.,  C. C. Moallemi, and M. G. Sotiropoulos,  2018,  Short-term trading skill: An analysis of investor heterogeneity and execution quality, \textit{Journal of Financial Markets}.
\bibitem{Frazzini} Frazzini, A., F. Israel, and T. J. Moskowitz, 2018, Trading Costs, Available at SSRN 3229719 .
\bibitem{BBLB} Bucci, F., M. Benzaquen, F. Lillo,  and J.-P. Bouchaud, 2018, Crossover from linear to square-Root market impact, Available at SSRN 3283687.
\end{thebibliography}
\end{document}